\documentclass[aps,pra,twocolumn,showpacs,floatfix,superscriptaddress]{revtex4}

\pdfoutput=1

\usepackage{amsmath}
\usepackage{amssymb}
\usepackage{graphicx}
\usepackage{epstopdf}

\newcommand{\bra}[1]{\left<#1\right|}
\newcommand{\ket}[1]{\left|#1\right>}

\begin{document}

\title{Layered Quantum Hall Insulators with Ultracold Atoms}
\author{A. Zamora}
\affiliation{ICFO-Institut de Ci\`encies Fot\`oniques, Mediterranean Techonoly Park, 08860 Castelldefels (Barcelona), Spain}
\author{G. Szirmai}
\affiliation{ICFO-Institut de Ci\`encies Fot\`oniques, Mediterranean Techonoly Park, 08860 Castelldefels (Barcelona), Spain}
\affiliation{Research Institute for Solid State Physics and Optics, H-1525 Budapest P.O. Box 49, Hungary}
\author{M. Lewenstein}
\affiliation{ICFO-Institut de Ci\`encies Fot\`oniques, Mediterranean Techonoly Park, 08860 Castelldefels (Barcelona), Spain}
\affiliation{ICREA-Instituci\'o Catalana de Recerca i Estudis Avan\c cats, Lluis Companys 23, 08010 Barcelona, Spain}

\begin{abstract}
We consider a generalization of the 2-dimensional (2D) quantum-Hall insulator to a non-compact, non-Abelian gauge group, the Heisenberg-Weyl group. We show that this kind of insulator is actually a layered 3D insulator with nontrivial topology. We further show that nontrivial combinations of quantized transverse conductivities can be engineered with the help of a staggered potential. We investigate the robustness and topological nature of this conductivity and connect it to the surface modes of the system. We also propose a simple experimental realization with ultracold atoms in 3D confined to a 2D square lattice with the third dimension being mapped to a gauge coordinate.
\end{abstract}


\date{\today}
\maketitle

\section{Introduction}
In recent years there has been a huge interest in topological states of matter and related external gauge fields realized with ultracold atoms on a 2-dimensional optical lattice. Historically the way has been paved by the study of rotating optical lattices, where the centrifugal force, appearing in a frame co-rotating with the lattice, acts as an artificial magnetic field (see Ref. \cite{bloch08a} for a review). To reach topologically nontrivial phases, however, needs a large magnetic field, which falls outside of the reach of rotating lattice experiments. Later a scheme has been proposed theoretically by Jaksch and Zoller \cite{jaksch2003a}, in which a strong artificial magnetic field can be created with the help of stationary optical forces, and the Hofstadter butterfly can be observed. Several modifications of this work appeared recently to overcome some difficulties \cite{gerbier2010a} or to generalize it to non-Abelian gauge potentials (SU(2)) \cite{osterloh05a,ruseckas05,jacob07a} giving rise to energy spectra with peculiar self-similarities beyond the Hofstadter butterfly \cite{osterloh05a,goldman09b} or to topological transport properties \cite{jacob08a,goldman09b,goldman09a,stanescu09a,stanescu10a,goldman10a}. Very recently Abelian gauge fields have been realized experimentally as well \cite{lin09a,lin09b}.

Probably the simplest paradigmatic example of topological transport is the integer quantum Hall (IQH) effect \cite{klitzing80a} taking place at low enough temperatures. By varying the magnetic field perpendicular to a 2-dimensional semiconductor, which also has an electric potential difference between two opposite edges, one can notice a transverse current. The transverse conductivity (TC) takes only quantized values and is proportional to the winding number of the Berry's curvature of the many-body wave function \cite{thouless82a}. Soon after the topological considerations of two-dimensional insulator phases the theory has been generalized to three dimensions where the TC is a tensor \cite{montambaux90a,kohmoto92a}. However, in three dimensions one usually suffers from the collapse of the energy gaps and the system remains an insulator only in special situations. Another generalization of the IQH effect to higher degrees of freedom is to stack more layers of more or less independent quantum Hall insulators (QHI) on top of each other. Since inter-layer tunneling is small well developed gaps remain. A lot of work has been devoted to study the effects of bi- or multi-layer structures on the quantum Hall effect, especially the fractional one \cite{wenbook,ezawabook}. In bilayer graphene the TC plateaus were studied and a new type of IQH effect was found with a zero-level anomaly interpreted by a $2 \pi$ Berry's phase of the charge carriers \cite{novoselov06a}.

Despite the above successes of realizing synthetic gauge fields with ultracold atoms, non-Abelian gauge groups have not yet been achieved experimentally. Here we propose a scheme where the difficulty arising from the multicomponent nature of particles sensitive to a non-Abelian gauge field is overcome quite naturally by mapping the gauge index to an external coordinate, thus rendering the realization of a non-Abelian external gauge field experimentally easier. The external coordinate is the position in a perpendicular direction. Depending on whether the extension of the system is finite or infinite in this direction, one can realize compact, or non-compact external gauge groups, respectively. In this paper we consider the non-compact case. We propose therefore an experimental scheme with single component ultracold atoms in a 3D optical lattice with cubic geometry but anisotropic hoppings to mimic the behavior of a 2D system but with (possibly) infinitely many internal states. 

Until now not much interest has been directed towards non-compact gauge groups, which is mainly rooted in  that these gauge groups have ill defined invariants (such as Wilson loops) and therefore their dynamics cannot be formulated in a gauge invariant manner. However, as external fields they can induce topological phases with peculiar transport properties. The question is what kind of physics does it yield, and what can be their physical realization? This paper answers positively to both questions. Here we study the effects of the simplest non-compact gauge group, namely the Heisenberg-Weyl group (HWG), which is generated by two elements, say $\hat{z}$ and $\hat{p}$, with the canonical commutation relation $[\hat{z},\hat{p}]=i$. As a consequence of non-compactness this commutator cannot be represented in finite dimensions. We show that in spite of the above difficulties this gauge group can be realized relatively simply with today's technology of ultracold fermions and optical lattices by identifying the role of the HWG in layered 3-dimensional lattice systems. We provide the phase diagram of the system. We also show that the multilayer structure provides further prospects beyond the zero level anomaly, such as a significant tunability of the positions and strengths of the quantum Hall plateaus. In a possible interferometric application such a strong transverse conductivity can, in principle, enhace the precision of the measurement.

The paper is organized as follows. In Sec. II we start our discussion with the 2D Hamiltonian describing non-interacting fermions in the presence of a general, possibly non-Abelian an non-compact, gauge field. This is the Hamiltonian we would like to simulate in ultracold atom experiments. We give a mapping of the gauge coordinate to the external $z$-direction, thus extending the spatial dimension by 1, and simultaneously reducing the wave function to a single component one. This mapping is the key point for experimental realization. We analyze the transport properties specific to the vector potential of our choice and relate the results to those of the IQH effect. In Sec. III we show that with the addition of a staggered scalar potential the transport properties can be modified drastically and give the phase diagram of the topologically insulating states. In Sec. IV we study the effects of some possible imperfections in the experimental applications, namely some small but nonzero inter-layer tunnelings and discuss the robustness of the phase diagram with respect to these imperfections. Sec. V is devoted to the experimental proposal and Sec. VI for summary and discussions.

\section{Lattice Hamiltonian in an external gauge field}
Our goal is to study topological transport properties of fermions confined in 2D and immersed to an external magnetic field generated by non-Abelian and non-compact gauge fields. To make the presentation more transparent and to show how a dimensional extension can be carried over we start our discussion with the 2D Hamiltonian of multicomponent fermions inside a periodic potential. We suppose that: 1) the fermions are non-interacting, which in experiments can be achieved e.g by Fesbach resonances; 2)  the optical lattice is sufficiently deep, therefore a tight-binding model can be used. The tight-binding Hamiltonian in second quantized form reads as 
\begin{multline}
\label{eq:ham2d}
H_{2d}=t\sum_{m,n,\sigma,\sigma'}\Big(U^x_{\sigma,\sigma'}(m,n)\,c^\dagger_{m,n,\sigma}c_{m+1,n,\sigma'}\\
+U^y_{\sigma,\sigma'}(m,n)\,c^\dagger_{m,n,\sigma}c_{m,n+1,\sigma'}+\mathrm{H.c.}\Big),
\end{multline}
where $c_{m,n,\sigma}$ is the fermion annihilation operator at position $\vec{r}=m\,b\,\vec{e}_x+n\,b\,\vec{e}_y$ ($b$ is the lattice spacing), and with gauge coordinate (internal state) $\sigma$. The hopping amplitude is given by $t$. The position dependent unitary matrices $U^x(m,n)$ and $U^y(m,n)$ are the lattice counterparts of the components of the vector potential $\vec{A}(\vec{r})$ of the continuum system, and are obtained by the Peierls substitution
\begin{subequations}
\begin{align}
U^x_{\sigma,\sigma'}(m,n)&=\Bigg[\exp\bigg(i\int_{\vec{r}_{m+1,n}}^{\vec{r}_{m,n}} \vec{A}(\vec{r})d\vec{r}\bigg)\Bigg]_{\sigma,\sigma'},\\
U^y_{\sigma,\sigma'}(m,n)&=\Bigg[\exp\bigg(i\int_{\vec{r}_{m,n+1}}^{\vec{r}_{m,n}} \vec{A}(\vec{r})d\vec{r}\bigg)\Bigg]_{\sigma,\sigma'}.
\end{align} 
\end{subequations}
In the case of standard electrodynamics, where the gauge group is Abelian, the tunneling operators are ordinary site dependent phase factors. When the gauge group is non-Abelian but compact they become unitary matrices with finite dimension. Here we choose them to be elements of the HWG, which is non-compact, and does not have finite dimensional representation, therefore the tunneling matrices become infinite dimensional. Here we consider the following choice
\begin{subequations}
\label{eqs:vectpot}
\begin{align}
U^x(m,n)&=e^{i\alpha \,\hat{p}},\\
U^y(m,n)&=e^{2\pi i(\beta \,m + \gamma \,\hat{z})}.
\end{align}
\end{subequations}
The operators $\hat{z}$ and $\hat{p}$ obey the commutation relations mentioned in the introduction, and they act on the internal space by $\hat{z}\,c_{m,n,\sigma}=\sigma\,c_{m,n,\sigma}$, and $e^{i\alpha \hat{p}}\,c_{m,n,\sigma}=c_{m,n,\sigma+\alpha}$, i.e. $\hat{z}$ is a diagonal operator (the position), while $\hat{p}$ is the generator of translations in the internal space. The $x-y$ position dependence of the vector potential has been chosen here similarly to the Landau gauge, which in the case of U(1) electrodynamics gives a homogeneous magnetic field important for the quantum Hall effect. The numeric parameters $\alpha$, $\beta$, and $\gamma$ are real numbers, which can be controlled experimentally and are measuring the flux penetrating each plaquette.

\begin{figure}[bt]
\begin{center}
\includegraphics[scale=0.75]{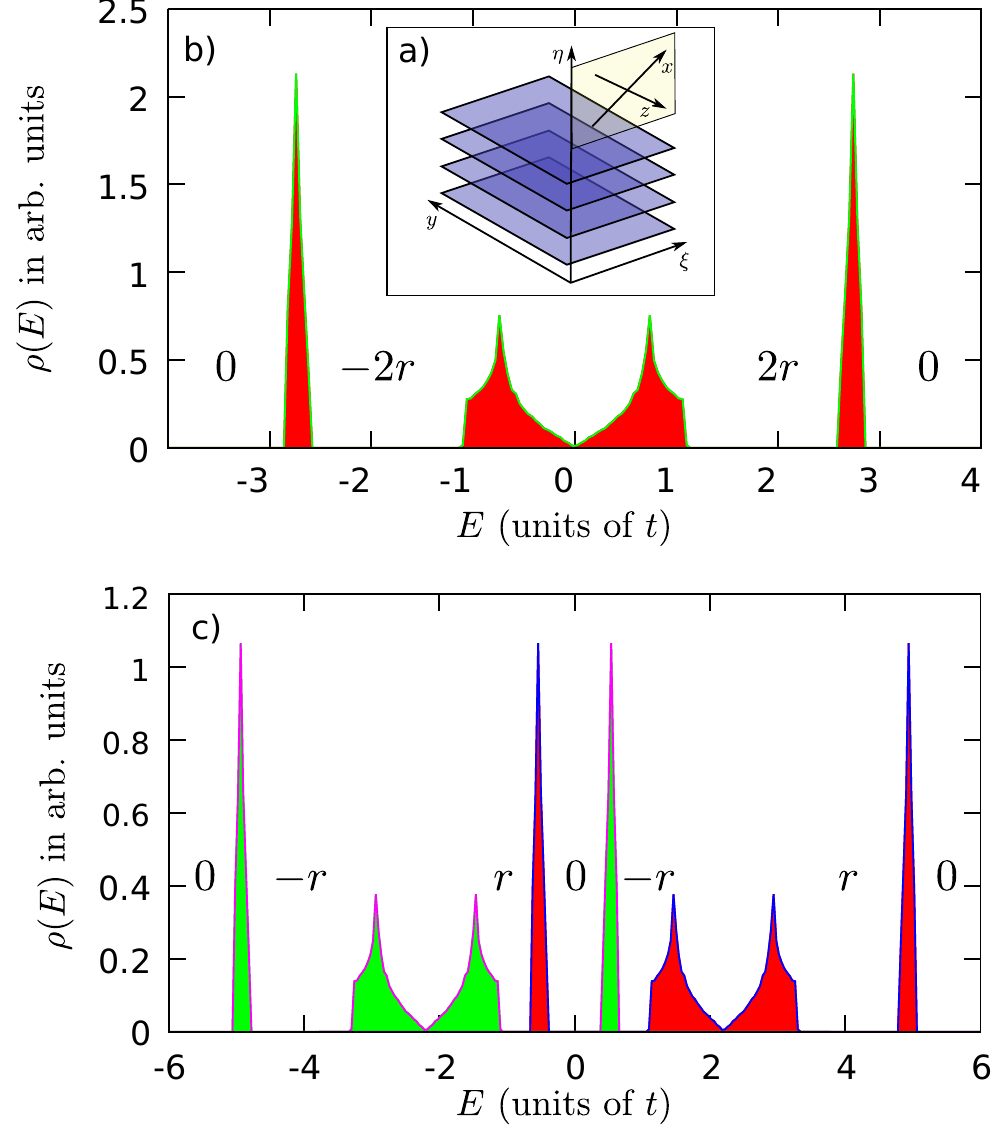}
  \caption{(color online) a) The schematic illustration of the new coordinates, the system splits into independent quantum Hall layers. b)-c) The DOS for $\beta'=1/4$ as a function
of the energy. In subfigure a) each different $\eta$ plane contributes equally to the DOS. In b) the value of the staggered potential in Eq. \eqref{eq:hamwithstag} is chosen to be $\lambda  = 2.2 t$. The net TCs are shown on the figure for each band gap.}
\label{fig:iqh}
\end{center}
\end{figure}

Let us consider that the gauge coordinate $\sigma$ is actually the real external $z$ direction. Let us further assume, that the $z$ direction also supports a sufficiently deep optical lattice with lattice spacing $b'$. So the 2-dimensional problem with infinitely many internal states is mapped to a true 3-dimensional problem, where the position is given by $\vec{r}=m\,b\,\vec{e}_x+n\,b\,\vec{e}_y+\sigma\,b'\,\vec{e}_z$. Since a particle can hop only from site to site -- even in the $z$ direction --, therefore $\alpha$ has to be integer times the lattice spacing. By considering the simplest choice, $\alpha=b'$, then $U^x(m,n)$ is just the translation operator in the $z$ direction by one lattice site. The first term in the lattice Hamiltonian \eqref{eq:ham2d} therefore takes the form $U^x(m,n)c^\dagger_{m,n,\sigma}c_{m+1,n,\sigma'}=c^\dagger_{m+1,n,\sigma-1}c_{m+1,n,\sigma}$. Accordingly the tunneling processes in the $xz$ plane are special; if the particle tunnels in the $x$ direction by one to the left (right), it has to tunnel one position in the $z$ direction up (down) too. Hence it is convenient to introduce new coordinates: $\xi=(m-\sigma)/2$, and $\eta=m+\sigma$, so the Hamiltonian takes the following form:
\begin{equation}
\label{eq:hamplanes}
H_{\text{H}}=t\sum_{\xi,\eta,n}\Big(c^\dagger_{\xi,\eta,n}c_{\xi+1,\eta,n}
+e^{i\theta_y}\,c^\dagger_{\xi,\eta,n}c_{\xi,\eta,n+1}+\mathrm{H.c.}\Big).
\end{equation}
The phase factor acquired by tunneling along the $y$-direction reads in the new coordinates as
\begin{equation}
\label{eq:phaseplanes}
\theta_y=2 \pi \beta' \xi + \pi \gamma'\eta,
\end{equation}
with the definitions $\beta'\equiv\beta-\gamma$, and $\gamma'\equiv\beta+\gamma$. As there is no tunneling along the $\eta$-direction, Eqs. \eqref{eq:hamplanes} and \eqref{eq:phaseplanes} describe independent 2-dimensional IQH systems layered on top of each other. Each layer behaves similarly, except for $\eta$ even $\xi$ takes integer values and for $\eta$ odd $\xi$ takes half integer values. The tunneling phase Eq. \eqref{eq:phaseplanes} depends also on $\eta$ (the quantum number indexing the different planes).  Refer to Fig. \ref{fig:iqh} a) for illustration.

It is convenient to choose $\beta'$ as a rational number: $\beta'=p/q$ with $p$ and $q$ co-prime integers. In this case the Hamiltonian is invariant under translations in the $\xi$ coordinate by $q$ lattice positions, homogeneous in the $y$ direction, while $\eta$ is a conserved quantity. For illustration we have calculated the density of states (DOS) for $\beta'=1/4$ by direct diagonalization and plotted in Fig \ref{fig:iqh} b). As a consequence of the $q=4$ periodicity in the $\xi$ direction, the energy spectrum is composed into $q=4$ bands which are the lattice counterparts of the Landau levels of the continuum case. The central two bands shows the features of a Dirac cone like touching. When the Fermi energy lies inside a band, the system is metallic and has a non-vanishing longitudinal conductivity. In contrast, when the Fermi energy lies inside a gap, the longitudinal conductivity is zero, and quantum Hall effect can take place: the transverse conductivity is integer times the conductivity quantum. Since the spectrum is independent of $\eta$, the effect of layering is just a degeneracy in the spectrum by the number of planes. For these (longitudinally) insulating phases the transverse conductivity of each plane is given by the TKNN formula \cite{thouless82a}:
\begin{equation}
\label{eq:transcond}
\sigma_{\eta}=\frac{1}{2\pi}\int d^2 k\,\mathcal{F}_\eta,
\end{equation}
with $\mathcal{F}_\eta=\nabla_k\times \mathcal{A}_\eta$ the Berry's curvature, and $\mathcal{A}_\eta=i\sum_{r=1}^N\bra{u^{(r)}_{k_\xi,k_y}}\nabla_k\ket{u^{(r)}_{k_\xi,k_y}}$ is the Berry's connection calculated for the occupied one particle Bloch states $u^{(r)}_{k_\xi,k_y}$ at plane $\eta$. We have computed the transverse conductivity with the efficient algorithm of Fukui et al \cite{fukui05a}. The transverse conductivity takes only integer values since it is a topological invariant: the first Chern class of the U(N) principal bundle over the Brillouin-zone torus $(k_\xi,k_y)$. Furthermore at every $\eta$ plane the transverse conductivity can be only $0$, $1$ or $-1$, depending on whether the Fermi energy lies outside the band structure (in this case the lattice is either empty or fully filled and the conductivity is zero), or it lies inside a gap between the satellite bands and the central band (for negative energies the transverse conductivity is negative and for positive energies it is positive for our model and choice of parameters). Accordingly, when we add up all of the contributions to the TC from each of the planes we get zero if the Fermi energy lies outside the band structure, or we get $\sigma_{\perp}=\sum_\eta\sigma_\eta=\pm2r$, where $2r$ is the number of planes, what we choose to be even. We have also shown in Fig. \ref{fig:iqh} b) and c) the net TC of the band gaps.

\section{Effects of a staggered potential}
\begin{figure}[tb!]
\begin{center}
\includegraphics[width=.5\textwidth]{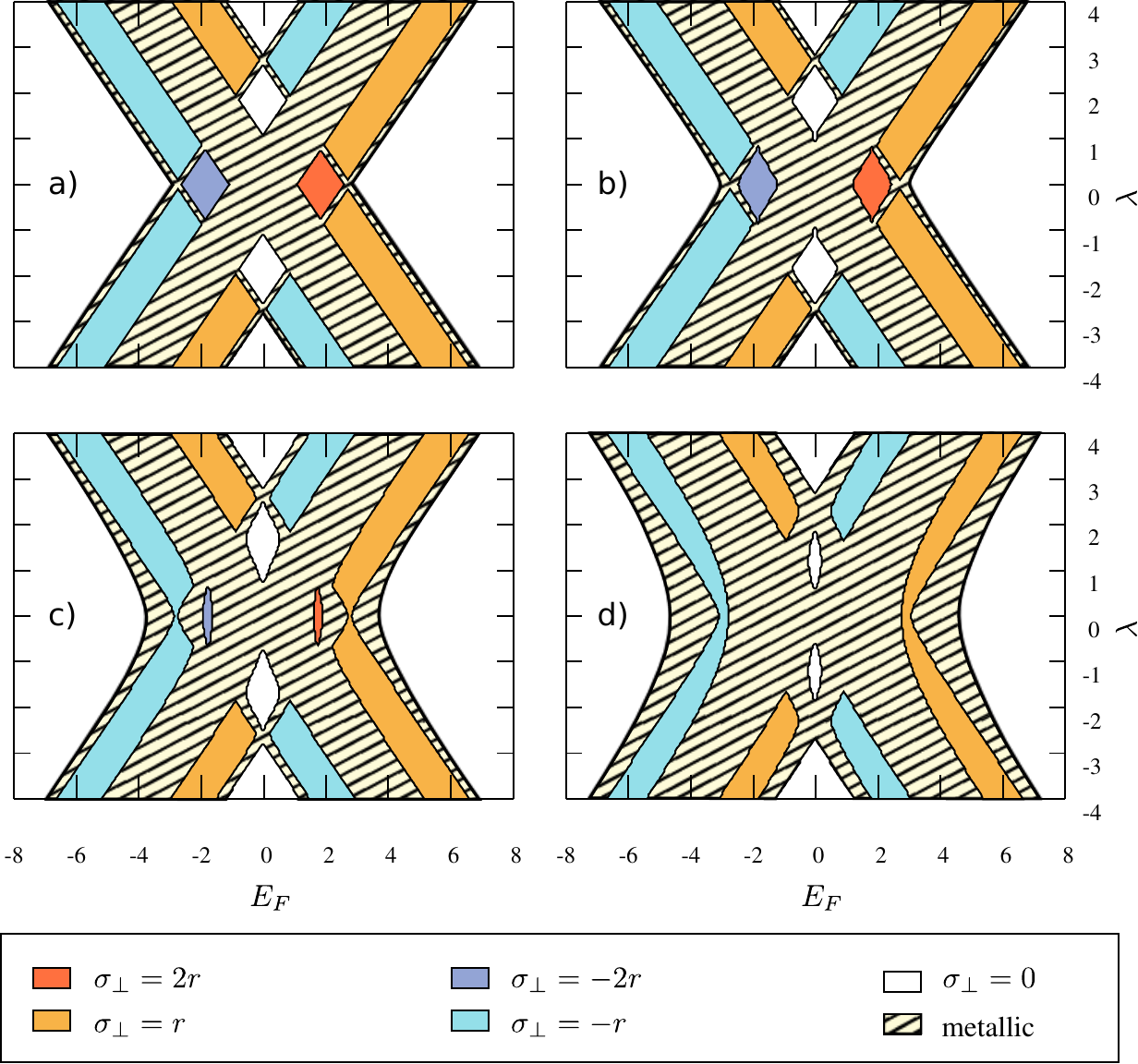}
\caption{(color online) The phase diagram of the system in the Fermi energy -- staggered potential plane (measured in the hopping strength). Subfigure a) corresponds to the Hamiltonian  \eqref{eq:hamwithstag}. Subfigure b)-d) correspond to Eq. \eqref{eq:hamfinal} with $\epsilon=0.1$ b), $\epsilon=0.5$ c), and $\epsilon=1.0$ d). The flux is taken to be $\beta'=1/4$.  The hatched regions represent areas where either the even or the odd planes are metallic.}
\label{fig:phdiaglam}
\end{center}
\end{figure}
The above combination rule of the TC can be controlled in a striking way by applying a staggered potential of strength $\lambda$, added to the Hamiltonian as
\begin{equation}
\label{eq:hamwithstag}
H_{\lambda}=H_{\text{H}}+\lambda\sum_{\xi,\eta,n}(-1)^\eta\,c^\dagger_{\xi,\eta,n}c_{\xi,\eta,n},
\end{equation}
which shifts the energy spectrum locally by $\pm\lambda$ for the even/odd planes. In Fig. \ref{fig:iqh} c) we have plotted the resulting DOS for $\lambda=2.2 t$. Note, that the original degeneracy is partially lifted by the $2\lambda$ energy difference of the planes with different lambda parities. One can imagine as we start to increase $\lambda$ from zero to a finite value that the original 3 bands start to widen and then separate, since the one particle energies on the even $\eta$ planes are getting bigger, while those on the odd $\eta$ planes are getting smaller.  The phase diagram in the $(E_F,\lambda)$ plane is plotted in Fig. \ref{fig:phdiaglam} a). One can get a grasp of the structure by imagining that the 3 energy bands of Fig. \ref{fig:iqh} b) get shifted by $\pm\lambda$ providing 3 tilted stripes of metallic behaviour for the even and and another 3 stripes (tilted to the other direction) for the odd planes. At the intersections of the stripes all of the planes are metallic at the same time. In the insulating regions, i.e. when the point $(E_F,\lambda)$ lies outside of the metallic stripes, the net TC is the sum of the contributions of the even and odd planes: $\sigma_{\perp}=\sigma_e+\sigma_o$. With the variation of $\lambda$ one can achieve that the Fermi energy falls into different band gaps for the even and odd planes, where $\sigma_e\neq\sigma_o$. In the $\beta'=1/4$ case the gaps are characterized by $\sigma_{e,o}=\lbrace0,-r,+r,0\rbrace$. Therefore one can have  $\sigma_\perp=-2 r$ if both set of planes have $\sigma_{e,o}=-r$, and $\sigma_\perp=+2r$ for $\sigma_{e,o}=+r$. Zero net TC is realized in two ways: either by $\sigma_{e,o}=0$, or by $\sigma_{e}=-\sigma_o$. In this latter case the even and odd planes have opposite transverse conductivities, which is similar to the spin quantum Hall effect (for a review see Ref. \cite{kane10a}). There is an important difference though: in our case the net TC is an integral number, not just a $Z_2$ invariant, the higher is its value the greater is the transverse current. If a quantum phase transition is induced either by the change of the Fermi energy or by the variation of $\lambda$, the TC changes at least by half of the number of planes, not just by one conductance quantum.

\begin{figure}[tb!]
\begin{center}
\includegraphics[scale=0.9]{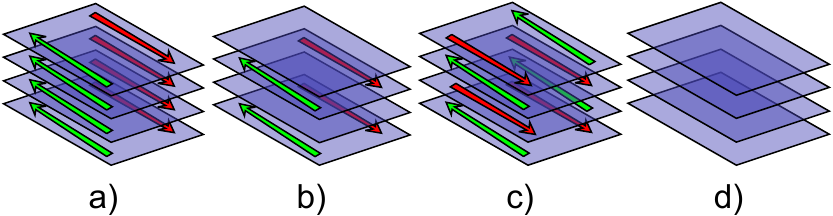}
\caption{(color online) The edge states. In situation a) all of the planes are in the same spectral gap, therefore the edge states of the different planes contribute equally to the transverse conductivity; in b) half of the planes is either metallic, totally empty or completely filled thus lacking the edge states, while the other half of the planes have a TC of 1; in c) the different parity planes have oppositely propagating edge states; in d) all of the planes are either metallic or trivial insulators, there are no edge states in this case.}
\label{fig:edgestates}
\end{center}
\end{figure}

Another physical picture can be assigned to the insulating regions of the phase diagram based on the bulk-boundary correspondence \cite{hatsugai93a}: by considering the $x$ direction finite, when the Fermi energy is inside a bulk gap with nonzero TC, edge states (responsible for the transverse current) traverse the energy spectrum between the two bulk bands surrounding the gap. In the $\beta'=1/4$ flux phase one can have a maximum of 1 pair of edge states per plane and the direction of their propagation is determined by the sign of the TC. Fig. \ref{fig:edgestates} shows a qualitative picture of the possible edge state configurations. In subfigure a) all planes are in the same QHI phase and the transverse currents carried by the edge states add together 'constructively'. Such a situation corresponds to the two diamond shaped regions with $\sigma_\perp=\pm2r$ of Fig. \ref{fig:phdiaglam} a). In subfigure \ref{fig:edgestates} b) every second plane is in a QHI phase while the other half of the planes are either in a metallic or in a normal insulator phase. Therefore $\sigma_\perp=\pm r$ and the corresponding regions in the phase diagram are those marked out by the IQH stripes (except for their intersecting diamond shape regions). The situation in Fig. \ref{fig:edgestates} c), where the edge states of the neighbouring planes are counter-propagating, is situated in the intersections of the oppositely transverse conducting IQH stripes of Fig. \ref{fig:phdiaglam} a). The net TC is zero and an analogy with quantum spin Hall insulators can be established \cite{kane10a}.

\begin{figure*}[t!]
\begin{center}
\includegraphics[width=\textwidth]{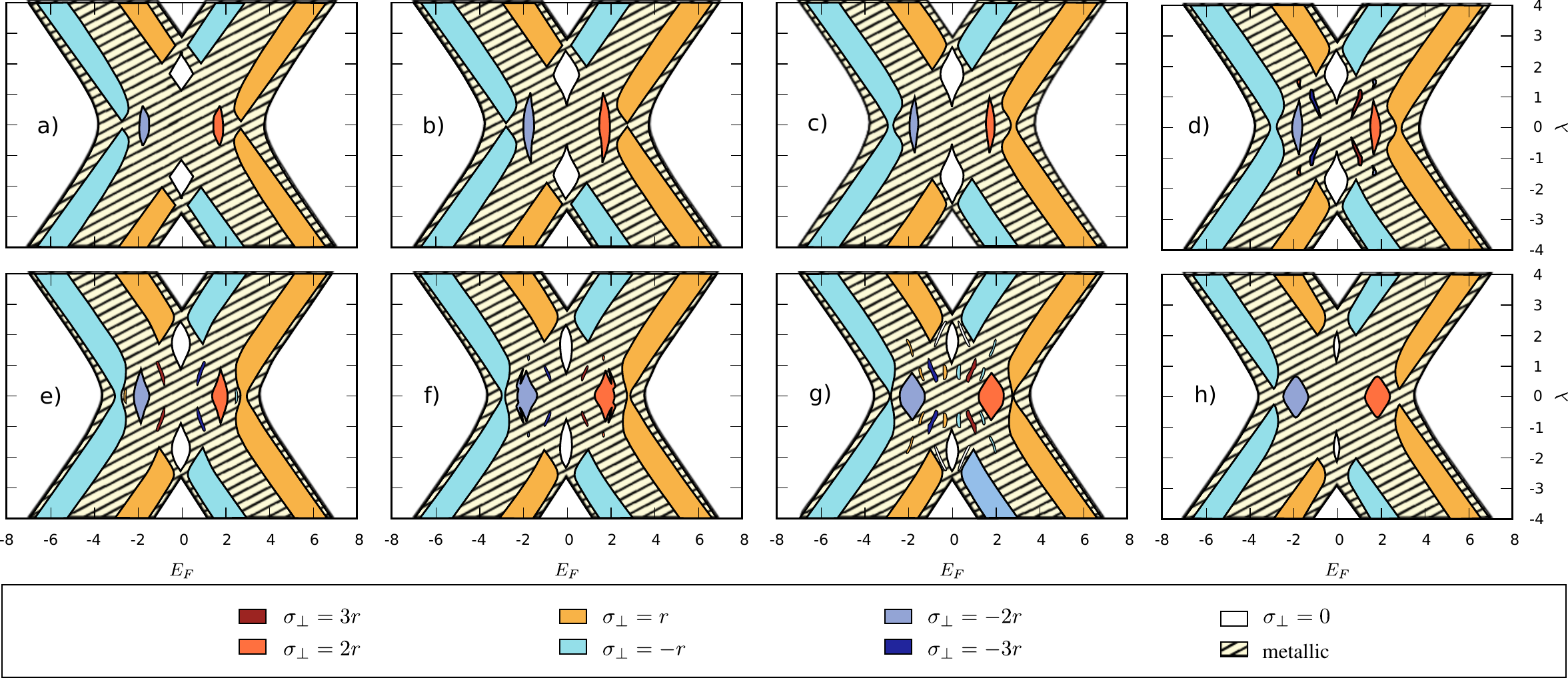}
\caption{(color online) The phase diagram for $\epsilon=0.5$, $\beta'=1/4$, $\gamma'=p'/q'$, with $q'=7$ and $p'=0,1,2,3,4,5,6,7$ from a) to h).}
\label{fig:etaphases}
\end{center}
\end{figure*}
\section{Interplane tunneling}
In experiments with ultracold atoms in anisotropic 3D lattices it is possible, that the inter-plane tunneling is not exactly zero, and therefore the HWG is realized perfectly. Therefore an analysis of the robustness of the accumulated TC is needed. One introduces tunneling between the different planes (in a controllable way) by adding an extra term to the Hamiltonian:
\begin{equation}
\label{eq:hamfinal}
H_{\lambda,\epsilon}=H_{\lambda}+\epsilon\sum_{\xi,\eta,n}\Big(c^\dagger_{\xi,\eta,n}c_{\xi,\eta+1,n}+\mathrm{H.c.}\Big),
\end{equation}
then $\eta$ will not be a good quantum number any longer, and consequently $\sigma_\eta$ of the individual planes looses its meaning. As an effect of the $\epsilon$ strength inter-plane hopping the eigenstates of the different $\eta$ planes get hybridized and the conducting regions get widened as shown in the phase diagram Fig \ref{fig:phdiaglam} b)-d). However, while $\epsilon$ is sufficiently small, the gaps persist and the net $\sigma_\perp$ keeps its significance. In this case the Brillouin zone is a 3-dimensional torus and one needs to make 2-dimensional slices to get topological invariants \cite{kohmoto92a}. In our case the only nonzero invariant is the one with slicing along a constant $k_\eta$ value. (For nonzero values of the other components of the conductivity tensor one needs first a gap closing and than a reopening.) As the inter-plane tunneling amplitude is raised the QHI regions shrink and give way to the metallic phase. There is a competition between the staggered potential strength $\lambda$ and the tunneling strength $\epsilon$. Therefore the high transverse conductivity QHI regions vanish first (around $\epsilon_c\approx0.8$) since they are located around $\lambda=0$.

When $\epsilon\neq0$ the edge currents are not protected against back-scattering by any discrete symmetry, in contrary to the case of the spin QHIs. If the state of the system is located in the $\sigma_\perp=0$ islands inside the metallic stripes, corresponding to the alternating transverse edge current setup of Fig. \ref{fig:edgestates} c), then inter-plane tunneling causes an effective back scattering, and the originally massless Dirac-like dispersions of the edge states develop mass gaps. The edge states of the other IQH regions ($\sigma_\perp\neq0$) retain their topologically protected nature and the massless touching of their Dirac cones.

So far we have not yet studied the consequences of varying the $\gamma'$ flux parameter of the vector potential \eqref{eq:phaseplanes}. While $\epsilon=0$ the distinct planes are independent and the inclusion of $\gamma'\neq0$ does not change the transport properties. In Fig. \ref{fig:phdiaglam}, the first panel corresponds essentially to uncoupled layers with constant synthetic Abelian magnetic field. This is because the HW gauge potential is restricted to $\epsilon=0$. In contrast, in Fig. \ref{fig:etaphases} the full non-Abelian nature  of the HW group is considerd by setting $\gamma'\neq 0$ together with $\epsilon\neq0$. The first panel again essentially reduces to the case of constant synthetic Abelian magnetic field, but for the coupled layers. The following panels illustrate then the effects of non-Abelian HW gauge fields on the conductivity in coupled layers. This effect  consist in appearance of novel insulating islands with transverse conductivities equal to $\pm 3 r$ in these new regions, which are bigger then those for $\gamma'=0$. It constitutes one of the improtant results of this paper. 

\section{Experiments}
The experimental realization of the HWG is straightforward. According to Eq. \eqref{eq:hamplanes} one needs to create a 3-dimensional optical lattice with cubic geometry in 2-dimensions ($\xi-y$ planes) and an AB type layering in the 3rd ($\eta$) direction. The gauge potential \eqref{eq:phaseplanes} is actually just a phase in this basis, and therefore it can be realized in the usual way \cite{jaksch2003a,gerbier2010a,lin09b}. One only needs a staggered potential along the 3rd ($\eta$) direction and a control over the tunneling amplitude along this direction. All of these elements are in the reach of current experimental technologies. The detection of the transverse conductivity and the edge states is a currently running issue. There are 3 proposals what we are aware of: a) by changing the atomic isotope to a bosonic one, loading them into and finally imaging them in the edge states \cite{stanescu09a}, b) with the help of stimulated Raman scattering one can measure one particle excitations directly \cite{dao07a}, and c) with the help of Bragg scattering one can directly measure the dynamical structure factor and hence also the single particle excitations \cite{liu10a}. Another proposal for the detection of the edge states in the presence of a trapping potential consist of creating a sharp interface, e.g. by controlling the hopping in the $y$ direction and exciting the edge state channel by a focused laser beam \cite{goldman10a}.

\section{Summary and discussion}
In this paper we have shown that the problem of 2-dimensional lattice fermions in an external non-compact Heisenberg gauge group is equivalent to an experimentally realizable, layered 3-dimensional Abelian quantum Hall system. Thus we have given a theoretical proposal for realizing non-Abelian and non-compact gauge groups with ultracold atoms in an optical lattice with phase imprinting techniques only. We have considered its advantages in engineering different QHI phases by varying the gauge potential and another external staggered potential. As an example we have calculated the phase diagram for the Abelian flux of $1/4$ per plaquette and related the bulk topological properties to the edge states of open boundary conditions. We have also shown that by varying the other flux parameter one can achieve a non-Abelian character when the layers are coupled sufficiently strongly. Layered fermion systems are encountered in many corners of condensed matter physics and our motivation was to find new insulating phases with nontrivial topologies to get a better understanding on the possible complicated phase diagrams found in such systems. In this respect our analysis opens a viewpoint on layered systems, where the layering can be understood as an additional gauge potential on a system with reduced dimensions.
Concerning practical points of view, the robustness of the topological quantization of measurable quantities can lead to metronomy standards or to applications in quantum information processing.

\section{Acknowledgements}
We acknowledge funding from the Spanish MEC projects TOQATA (FIS2008-00784), QOIT (Consolider Ingenio 2010), ERC Advanced Grant QUAGATUA, EU STREP NAMEQUAM, NSF  (NF68736, T077629) and NORT (ERC\_HU\_09 OPTOMECH) of Hungary (G.Sz.), and Alexander von Humboldt Foundation (M.L.).


\end{document}